\input{aipcheck}
\documentclass[
    ,final            
  ]
  {aipproc}

\layoutstyle{8x11single}

\usepackage{graphicx,xcolor}

\newcommand{\beq}{\begin{equation}}
\newcommand{\eeq}{\end{equation}}
\newcommand{\bea}{\begin{eqnarray}}
\newcommand{\eea}{\end{eqnarray}}

\begin{document}

\title{Semihard  processes with BLM renormalization scale setting}


\author{Francesco Caporale}{address={Instituto de F\'isica Te\'orica UAM/CSIC,
Nicol\'as Cabrera 15 \\
and U. Aut\'onoma de Madrid, E-28049 Madrid, Spain}
}

\author{Dmitry Yu.~Ivanov}{address={Sobolev Institute of Mathematics 
and Novosibirsk State University, 630090 Novosibirsk}
}

\author{Beatrice Murdaca}{address={Dipartimento di Fisica, Universit\`a della 
Calabria, \\ and Istituto Nazionale di Fisica Nucleare, Gruppo collegato di
Cosenza, \\ Arcavacata di Rende, I-87036 Cosenza, Italy}
}

\author{Alessandro Papa}{address={Dipartimento di Fisica, Universit\`a della 
Calabria, \\ and Istituto Nazionale di Fisica Nucleare, Gruppo collegato di
Cosenza, \\ Arcavacata di Rende, I-87036 Cosenza, Italy}
}

\begin{abstract}
We apply the BLM scale setting procedure directly to amplitudes 
(cross sections) of several semihard processes. It is shown that, due to the 
presence of $\beta_0$-terms in the NLA results for the impact factors, 
the obtained optimal renormalization scale is not universal, but depends 
both on the energy and on the process in question. We illustrate this general 
conclusion considering the following semihard processes: 
(i) inclusive production of two forward high-$p_T$ jets separated by large 
interval in rapidity (Mueller-Navelet jets); (ii) high-energy behavior of 
the total cross section for highly virtual photons; (iii) forward amplitude 
of the production of two light vector mesons in the collision of two virtual 
photons.
\end{abstract}

\maketitle


\section{Introduction}
We discuss the application of the BFKL method~\cite{BFKL} to the description 
of semihard  processes, i.e. processes in the kinematic region 
where the energy variable $s$ is substantially larger than the hard scale 
$Q^2$, namely $s\gg Q^2\gg \Lambda^2_{\rm QCD}$, where $Q$ is the typical
transverse momentum and $\Lambda_{\rm QCD}$ is the QCD scale. 

The BFKL approach allows to resum systematically in all orders of 
perturbation series the terms enhanced by leading logarithms of energy,
$\alpha_s^n \ln^n(s/Q^2)$ (leading logarithmic approximation or LLA) and
sub-leading ones, $\alpha_s^{n+1} \ln^n(s/Q^2)$ (next-to-LLA or NLA). 
The practical application of BFKL to physical processes encounters serious 
difficulties due to large NLA corrections and big renormalization scale 
setting uncertainties. Due to that, one needs to apply some optimization method
to the QCD perturbative series. We adopt here the Brodsky-Lepage-Mackenzie (BLM)
approach~\cite{BLM} that eliminates the renormalization scale ambiguity by 
absorbing the non-conformal {$\beta_0$}-terms into the running coupling. 
In particular, we apply this optimization method to (i) the production of two 
forward high-$p_T$ jets separated by a large interval in rapidity 
(Mueller-Navelet jets~\cite{MN}), (ii) the total cross section for the 
collision of two off-shell photons with large virtualities and (iii) 
the forward amplitude for the electroproduction of two light vector mesons.

\section{BFKL amplitude and BLM scale setting}
Within the class of semihard processes, we are interested in the study of 
physical observables directly related to the forward amplitude. One of these
observables is the cross section itself that, due to the optical theorem,
can be written as $\sigma =\frac{{\rm Im}_s  {\cal A}}{s} $.

In the BFKL approach, both in the LLA and in the NLA, ${\rm Im}_s {\cal A}$ is 
given as the convolution of the Green's function ($G_{\omega}$) of two 
interacting Reggeized gluons with momenta $q_{1,2}$ and of the impact factors of 
the colliding particles ($\Phi_1$ and $\Phi_2$):
\begin{equation}
{\rm Im}_s \left( {\cal A} \right)=\frac{s}{(2\pi)^{2}}\int\frac{d^{2}\vec q_1}
{\vec q_1^{\,\, 2}}\Phi_1(\vec q_1,s_0)\int
\frac{d^{2}\vec q_2}{\vec q_2^{\,\,2}} \Phi_2(-\vec q_2,s_0)
\int\limits^{\delta +i\infty}_{\delta
-i\infty}\frac{d\omega}{2\pi i}\left(\frac{s}{s_0}\right)^\omega
G_\omega (\vec q_1, \vec q_2)\, ,
\end{equation}
where $s$ is the squared center-of-mass energy of the colliding particles, 
while $s_0$ is an artificial scale introduced in the BFKL approach to perform 
the Mellin transform from the $s$-space to the complex angular momentum plane.
Moving to the $(n,\nu)$-representation, which means using the basis of the
eigenfunctions of the leading-order (LO) BFKL kernel instead of that of 
transverse momenta, we get
\[
{\rm Im}_s
\left( {\cal A} \right)=\frac{s}{(2\pi)^2}\sum^{\infty}_{n=-\infty}
\int\limits^{\infty}_{-\infty} d\nu \left(\frac{s}{s_0}\right)^{\bar \alpha_s(\mu_R)
\chi(n,\nu)} \alpha_s^2(\mu_R) c_1(n,\nu)c_2(n,\nu)
\left\{1+\bar \alpha_s(\mu_R)\left(\frac{c^{(1)}_1(n,\nu)}{c_1(n,\nu)}
+\frac{c^{(1)}_2(n,\nu)}{c_2(n,\nu)}\right) \right.
\]
\begin{equation}
\label{ampl-ff}
\left.
+\bar \alpha^2_s(\mu_R)\ln\frac{s}{s_0}\left[\bar \chi(n,\nu)
+\frac{\beta_0}{8 N_c}\chi(n,\nu)\left(-\chi(n,\nu)+\frac{10}{3}
+2\ln \mu_R^2 +i\frac{d}{d\nu}\ln\frac{c_1(n,\nu)}{c_2(n,\nu)}\right)
\right]\right\} \;.
\end{equation}
The cross section could be represented, accordingly, as
\begin{equation}
\sigma= \frac{1}{\left(2\pi\right)^2}\left[ C_0+\sum_{n=1}^{\infty}2
\cos\left(n\phi\right)C_n\right]\, .
\end{equation}
Here $c_{i=1,2}(n,\nu)$ and $c_{i=1,2}^{(1)}(n,\nu)$ are the impact factors at 
the LO and next-to-LO (NLO) in the $(n,\nu)$-representation, respectively; 
$\chi(n,\nu)$ is the LO BFKL characteristic function and $\bar\chi(n,\nu)$ 
is related with the NLO correction to the BFKL kernel. For more details on the
derivation of Eq.~(\ref{ampl-ff}), see {\emph e.g.} Ref.~\cite{BLM_paper}.  
Note that $C_0$ is the total cross section, while the $C_n$'s with $n\neq 0$
are associated to azimuthal angle correlations, as in the case of 
Mueller-Navelet jets.

According to the BLM method, the renormalization scale $\mu_R$ in an 
the expression for a certain observable is chosen such that it makes
the $\beta_0$-dependent part vanish. For the observable $C_n$, originally
given in the $\overline{\rm MS}$ scheme, we first perform a finite 
renormalization transformation to the physical MOM scheme, which implies
the replacement
\begin{equation}
\label{scheme}
\alpha_s^{\overline{\rm MS}}=\alpha_s^{\rm MOM}\left(1+\frac{\alpha_s^{\rm MOM}}{\pi}
T \right)\;, \;\;\;\;\; T=T^{\beta}+T^{\rm conf}\;,
\end{equation}
\[
T^{\beta}\ =\ -\frac{\beta_0}{2}\left( 1+\frac{2}{3}I \right)\;,\;\;\;\;\;
T^{\rm conf}\ =\ \frac{C_A}{8}\left[ \frac{17}{2}I +\frac{3}{2}\left(I-1\right)
\xi+\left( 1-\frac{1}{3}I\right)\xi^2-\frac{1}{6}\xi^3 \right]\;,
\]
where $I=-2\int_0^1dx\frac{\ln\left(x\right)}{x^2-x+1}\simeq 2.3439$ and 
$\xi$ is a gauge parameter. We thus get
\[
C^{\rm MOM}_n
=\frac{1}{(2\pi)^2}\int\limits^{\infty}_{-\infty} d\nu \left(\frac{s}{s_0}
\right)^{\bar \alpha^{\rm MOM}_s(\mu_R)\chi(n,\nu)} \left(\alpha^{\rm MOM}_s (\mu_R)
\right)^2 c_1(n,\nu)c_2(n,\nu)
\]
\beq
\label{ampl-MOM}
\times\left[1+\bar \alpha^{\rm MOM}_s(\mu_R)\left\{\frac{\bar c^{(1)}_1(n,\nu)}
{c_1(n,\nu)}+\frac{\bar c^{(1)}_2(n,\nu)}{c_2(n,\nu)}+\frac{2T^{\rm conf}}{N_c}
+\frac{\tilde c_1^{(1)}}{c_1}+\frac{\tilde c_2^{(1)}}{c_2}
\right\}
\right.
\eeq
\[
+\left(\bar \alpha^{\rm MOM}_s(\mu_R)\right)^2\ln\frac{s}{s_0}
\left\{\!\bar\chi(n,\nu) +\frac{T^{\rm conf}}{N_c}\chi(n,\nu)
\left.
+\frac{\beta_0}{4 N_c}\chi(n,\nu)\left(\!
-\frac{\chi(n,\nu)}{2}+\frac{5}{3}+\ln \frac{\mu_R^2}{Q_1 Q_2} 
+f(\nu)-2\left( 1+\frac{2}{3}I \right)\right)\right\}\right] \,,
\]
where $\tilde c_{1,2}^{(1)}$ represent the part of the impact factor depending
on $\beta_0$, while $\bar c_{1,2}^{(1)}$ is the rest. In particular, the 
contribution to the NLO impact factor that is proportional to $\beta_0$ is 
universally expressed through the lowest order impact factors,
\beq
\frac{\tilde c_1^{(1)}}{c_1}+\frac{\tilde c_2^{(1)}}{c_2}\ 
=\ \frac{\beta_0}{2N_c}\left[ \frac{5}{3}+\ln \frac{\mu_R^2}{Q_1 Q_2} 
+f(\nu)\right]\;,
\eeq
where the function $f(\nu)$ depends on the process considered and is given by
\[
i\frac{d}{d\nu}\ln\left(\frac{c_1}{c_2}\right) \equiv 2\left[f(\nu)
-\ln\left(Q_1Q_2\right)\right]\;.
\]
Here $Q_{1,2}$ denotes the hard scale which enters impact factor $c_{1,2}$.

Now the optimal scale $\mu_R^{\rm BLM}$ is chosen such that it makes 
the expression proportional to $\beta_0$ vanish:
\[
C^{\beta}_n
=\frac{1}{(2\pi)^2}\int\limits^{\infty}_{-\infty} d\nu \left(\frac{s}{s_0}
\right)^{\bar \alpha^{\rm MOM}_s(\mu^{\rm BLM}_R)\chi(n,\nu)} \left(\alpha^{\rm MOM}_s 
(\mu^{\rm BLM}_R)\right)^3
\]
\beq
\label{c_nnnbeta}
\times c_1(n,\nu)c_2(n,\nu) \frac{\beta_0}{2 N_c} \left[\frac{5}{3}
+\ln \frac{(\mu^{\rm BLM}_R)^2}{Q_1 Q_2} +f(\nu)-2\left(1+\frac{2}{3}I \right)
\right.
\end{equation}
\[
\left.
+\bar \alpha^{\rm MOM}_s(\mu^{\rm BLM}_R)\ln\frac{s}{s_0} \frac{\chi(n,\nu)}{2}
\left(-\frac{\chi(n,\nu)}{2}+\frac{5}{3}+\ln \frac{(\mu^{\rm BLM}_R)^2}{Q_1 Q_2}
+f(\nu)-2\left( 1+\frac{2}{3}I \right)\right)\right]=0 \;.
\]
Unfortunately, Eq.~(\ref{c_nnnbeta}) can be solved only numerically. 
Therefore here we will limit ourselves to approximate approaches to the BLM 
scale setting. We consider the BLM scale as a function of $\nu$ and choose 
it in order to put equal to zero either the terms in the second
line of Eq.~(\ref{c_nnnbeta}) (case $(a)$) or those in the third line
of the same equation (case $(b)$). This corresponds to the
vanishing of the $\beta_0$-terms originating from the NLO corrections of the 
impact factors or of the BFKL kernel, respectively. We have then the 
two following cases:
\bea
\label{casea}
\left(\mu_{R, a}^{\rm BLM}\right)^2 &=& 
Q_1Q_2\ \exp\left[2\left(1+\frac{2}{3}I\right)
-f\left(\nu\right)-\frac{5}{3}\right]\;, \\
\label{caseb}
\left(\mu_{R, b}^{\rm BLM}\right)^2 &=&
Q_1Q_2\ \exp\left[2\left(1+\frac{2}{3}I\right)-f\left(\nu\right)-\frac{5}{3}
+\frac{1}{2}\chi\left(\nu,n\right)\right]\; .
\eea

\section{Application to some semihard processes}
Now we test the two solutions, Eqs.~(\ref{casea}) and (\ref{caseb}), on 
different semihard processes: (i) the inclusive production of Mueller-Navelet 
jets, (ii) the collision of two highly-virtual photons at large energies
and (iii) the production of two light vector mesons in the collision of two 
virtual photons, $\gamma^*\gamma^*\to V_1V_2$.

For these particular processes we have $f(\nu)=0$ in the first two cases, 
while $f(\nu)\ =\ \psi\left( 3+2i\nu\right)
+\psi\left( 3-2i\nu\right)-\psi\left( \frac{3}{2}+i\nu\right)
-\psi\left( \frac{3}{2}-i\nu\right)$ for the case of 
meson pair production.

Some of our results are reported in Fig.~\ref{plots}. In particular we show
the forward amplitude for the $\gamma^*\gamma^*\to V_1V_2$ process and the 
cross section of the Mueller-Navelet jets production process as a function of 
the rapidity interval between the produced pair of mesons or jets. For the
explicit definition of the considered quantities, see~\cite{Ivanov:2005gn} 
and~\cite{jet_paper}, correspondingly.
  
Other results for Mueller-Navelet jets, including moments of azimuthal 
angular correlations between jets, can be found in Fig.~3 of~\cite{jet_paper}.

Our results for the application of BLM method to the high-energy behavior 
of the total cross section for highly virtual photons is given instead 
in Fig.~6 of~\cite{photon_paper} (see also~\cite{Caporale:2008is}).

\section{Discussion}

\begin{figure}[t]
\begin{minipage}{0.50\textwidth}
\centering
\includegraphics[scale=0.45]{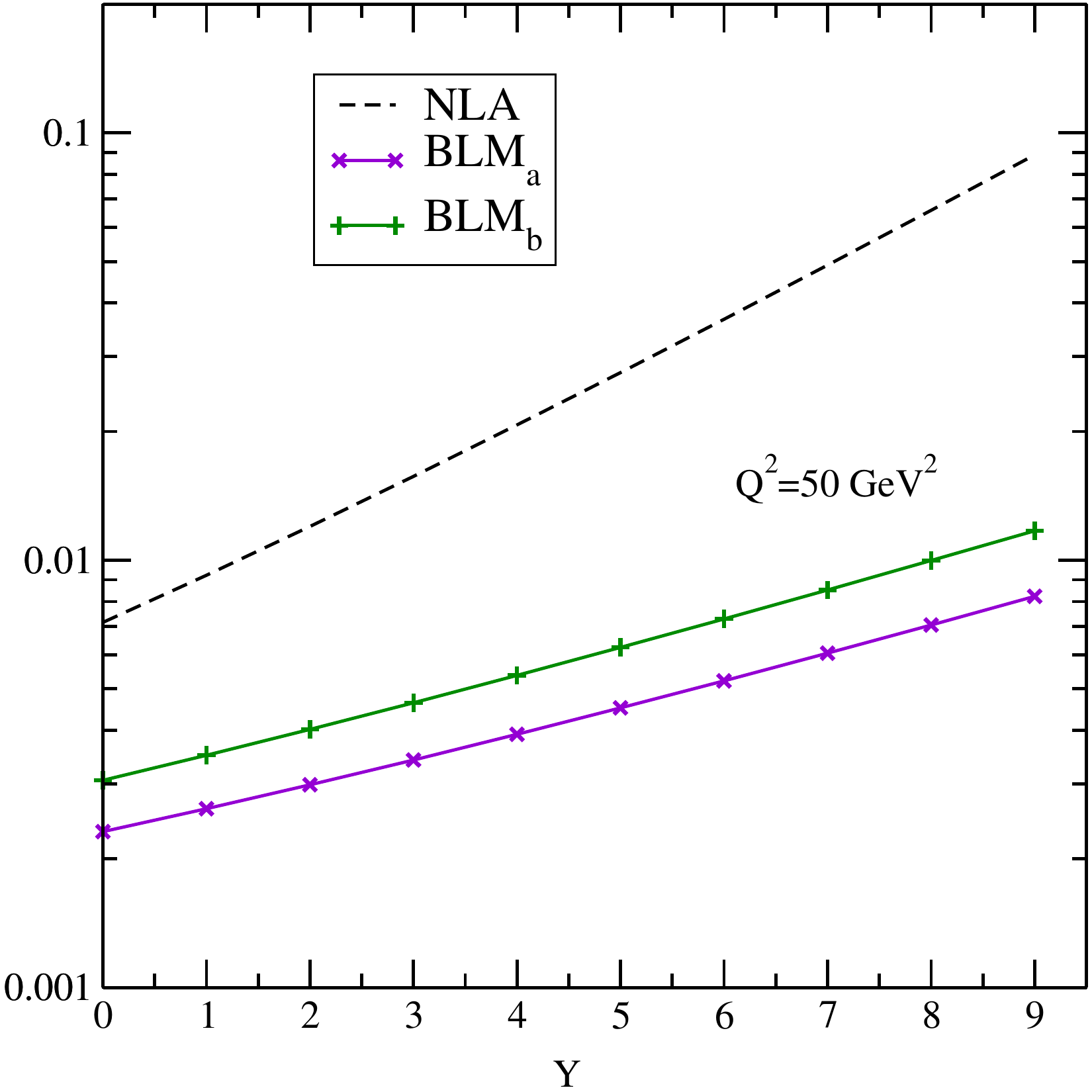}
\end{minipage}
\begin{minipage}{0.50\textwidth}
\centering
\includegraphics[scale=0.45]{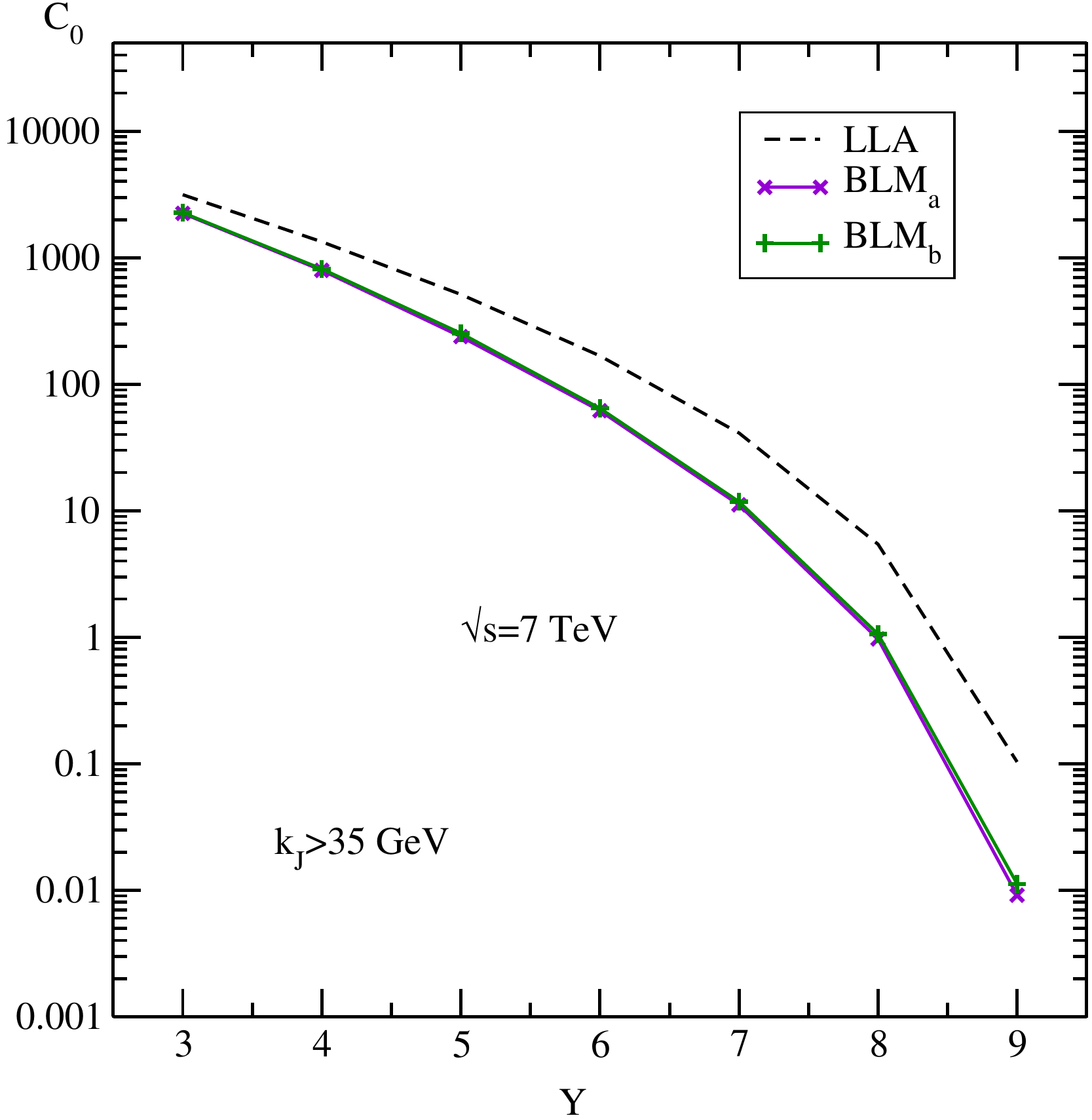}
\end{minipage}
\caption{Forward amplitude for the process $\gamma* \gamma* \to V_1 V_2 $ 
(left) and cross section of $p+p \to \textup{jet} + \textup{jet} + X$ 
(right).}
\label{plots}
\end{figure}

We note that in all cases the NLA corrections are very large and important. 
To illustrate this statement,  we show in Fig.~\ref{plots} also the LLA 
contributions, calculated for natural ({\it i.e.} non-optimal) choices
of the energy and renormalization scales. Due to this fact, NLA predictions 
for semihard processes depend substantially on the choice of energy scale 
$s_0$ and on the value of renormalization scale $\mu_R$. We adopted here the
BLM procedure to fix $\mu_R$.  

The application of the BLM approach to semihard reactions was pioneered 
in~\cite{Brodsky:1998kn}, where the method was applied to calculate the 
intercept of the BFKL Pomeron in the NLO. Later on, in~\cite{Brodsky:2002ka} 
the total cross section of $\gamma^*\gamma^*$ interactions at high energy was 
studied. Recently it was shown in~\cite{Ducloue:2013bva} that the application 
of the BLM method leads to a successful phenomenology of Mueller-Navelet jet 
production at LHC. 

One should mention, however, that the implementation of the BLM method 
in~\cite{Brodsky:2002ka,Ducloue:2013bva} was approximate, because it did not 
take into account the $\beta_0$-dependent parts of the NLA forward amplitudes 
originating from the NLO impact factors. Instead, only contributions coming 
from the NLO corrections to the BFKL kernel were considered, which 
corresponds to our case $(b)$, given in Eq.~(\ref{caseb}). Our case $(a)$, 
see~(\ref{casea}), means instead the approximation when one takes into 
account the $\beta_0$-terms coming only from the impact factors and 
neglects those originating from the kernel. 

Note that an explicit solution of the BLM scale setting 
equation~(\ref{c_nnnbeta}) leads to optimal scale values which depend on the 
process energy, $s$. To some extent, the explicit solution 
of~(\ref{c_nnnbeta}) gives a scale that interpolates between the approximate 
ones presented in~(\ref{casea}) and~(\ref{caseb}). Note also the BLM scale 
derived from~(\ref{c_nnnbeta}) is not universal, but depends on the process 
in question, through the form of the LO impact factors, $c_{1,2}(n,\nu)$.  
 
We observe that the difference of the two predictions for the 
Mueller-Navelet jet production cross section derived here using the 
approximate choices $(a)$ and $(b)$ of the BLM scale is not big. In our 
forthcoming paper~\cite{BLM_paper} we will present results for this and other 
Mueller-Navelet observables obtained also with the ``exact'' determination of 
BLM scale, as defined by Eq.~(\ref{c_nnnbeta}). We will study also the 
dependence of the predictions obtained with BLM method on the variation 
of the energy scale parameter $s_0$.

\begin{theacknowledgments}
The work of D.I. was also supported in part by the grant RFBR-13-02-00695-a.
The work of B.M. was supported by the European Commission, European Social
Fund and Calabria Region, that disclaim any liability for the use that can be
done of the information provided in this paper.
\end{theacknowledgments}

\bibliographystyle{aipproc}   

\end{document}